\documentclass[twocolumn,showpacs,preprintnumbers,pre,floatfix]{revtex4}
\usepackage{graphicx}
\usepackage{color}
\usepackage{epsfig}
\usepackage{amsmath}
\usepackage{amssymb}
\usepackage{verbatim}
\usepackage{graphics}
\usepackage{natbib}

\usepackage{float}
\floatstyle{ruled}

\usepackage{pstricks}
\usepackage{eepic}

\begin{document}

\title{Jamming of frictional spheres and random loose packing}

\author{Leonardo E. Silbert\footnote{lsilbert@physics.siu.edu}}

\affiliation{Department of Physics, Southern Illinois University, Carbondale,
  IL, USA.}

\begin{abstract}

  The role of friction coefficient, $\mu$, on the jamming properties of
  disordered, particle packings is studied using computer simulations.
  Compressed, soft-sphere packings are brought towards the jamming transition
  - the point where a packing loses mechanical stability - by decreasing the
  packing fraction. The values of the packing fraction at the jamming
  transition, $\phi^{\mu}_{c}$, gradually decrease from the random close
  packing point for zero friction, to a value coincident with random loose
  packing as the friction coefficient is increased over several orders of
  magnitude.  This is accompanied by a decrease in the coordination number at
  the jamming transition, $z^{\mu}_{c}$, which varies from approximately six
  to four with increasing friction. Universal power law scaling is observed in
  the pressure and coordination number as a function of distance from the
  generalised, friction-dependent jamming point.  Various power laws are also
  reported between the $\phi^{\mu}_{\rm c}$, $z^{\mu}_{\rm c}$, and
  $\mu$. Dependence on preparation history of the packings is also
  investigated.

\end{abstract}
 
\maketitle

\section*{Introduction}

Granular packings can exist over a range of densities generally depending on
the generation protocol and the nature of the grain-grain interactions. For
{\em frictionless} spheres, where the particle friction coefficient $\mu =0$,
{\em random close packing} (rcp), or the maximally random jammed state,
describes the densest possible packing of dry, cohesionless, spheres whose
structure contains no long-range order
\cite{bernal4,scott1,berryman1,torquato1,makse1,ohern3,torquato5,leo17}.
Random close packing is well-defined in the sense that it represents a
reproducible packing state as observed in numerous experiments and simulations
of frictionless sphere packings of hard particles that interact primarily
through excluded volume effects \cite{liu3}. All such studies agree that
random close packing occurs at a packing fraction, $\phi_{\rm rcp}$ $\approx
0.64$. This is in contrast to {\it frictional} packings, $\mu>0$, which can
exist over a substantial range in packing fraction from $\phi_{\rm rcp}$ all
the way down to $\phi_{\rm rlp} \approx 0.55$, the value often quoted as {\it
  random loose packing} \cite{liniger1,schroter3}. Although, in reality, it is
actually quite difficult to generate loose packed states. Consequently, random
loose packing is a much less well-developed concept than random close
packing. The strong history dependence and characterisation of frictional
packings persists as an experimental issue thus opening the door for
simulations to shed light on the nature of random loose packing of frictional
particles.

Frictionless sphere-packings and the rcp state have received much focus in
terms of {\it jamming} - the transition from a jammed, rigid, solid-like
state, with finite shear and bulk moduli, to an unjammed, liquid-like state
\cite{liu2,coniglio1,ohern2,ohern3}. The simplest system exhibiting such a
transition occurs in a packing of over-compressed, purely-repulsive, soft,
frictionless spheres, at zero temperature in the absence of gravity. As the
packing fraction, $\phi$, is decreased, the packing undergoes a transition at
$\phi_{\rm c} $, between jammed and unjammed phases that occurs abruptly at
$\phi_{\rm c} = \phi_{\rm rcp}$. Approaching the jamming transition from
above, $\phi \rightarrow \phi_{\rm c}^{+}$, the average number of contacting
neighbours per particle - the coordination number $z$ - approaches the minimal
value required for mechanical stability $z_{\rm iso}$, also known as
isostaticity. For frictionless spheres, $z_{\rm iso}^{\mu=0} = 6$ (in $3D$,
$=4$ in $2D$) \cite{alexander1,moukarzel1,roux1}. Thus, the interplay between
mechanical stability and maximally random \cite{torquato1}, in some sense,
provides a meaningful operational definition of the rcp state. Although it is
worthwhile to note that the particle positions are disordered - no long range
order exists - but they are not completely random as they must satisfy
mechanical equilibrium.

\begin{figure}[h]
  \includegraphics[width=7.5cm]{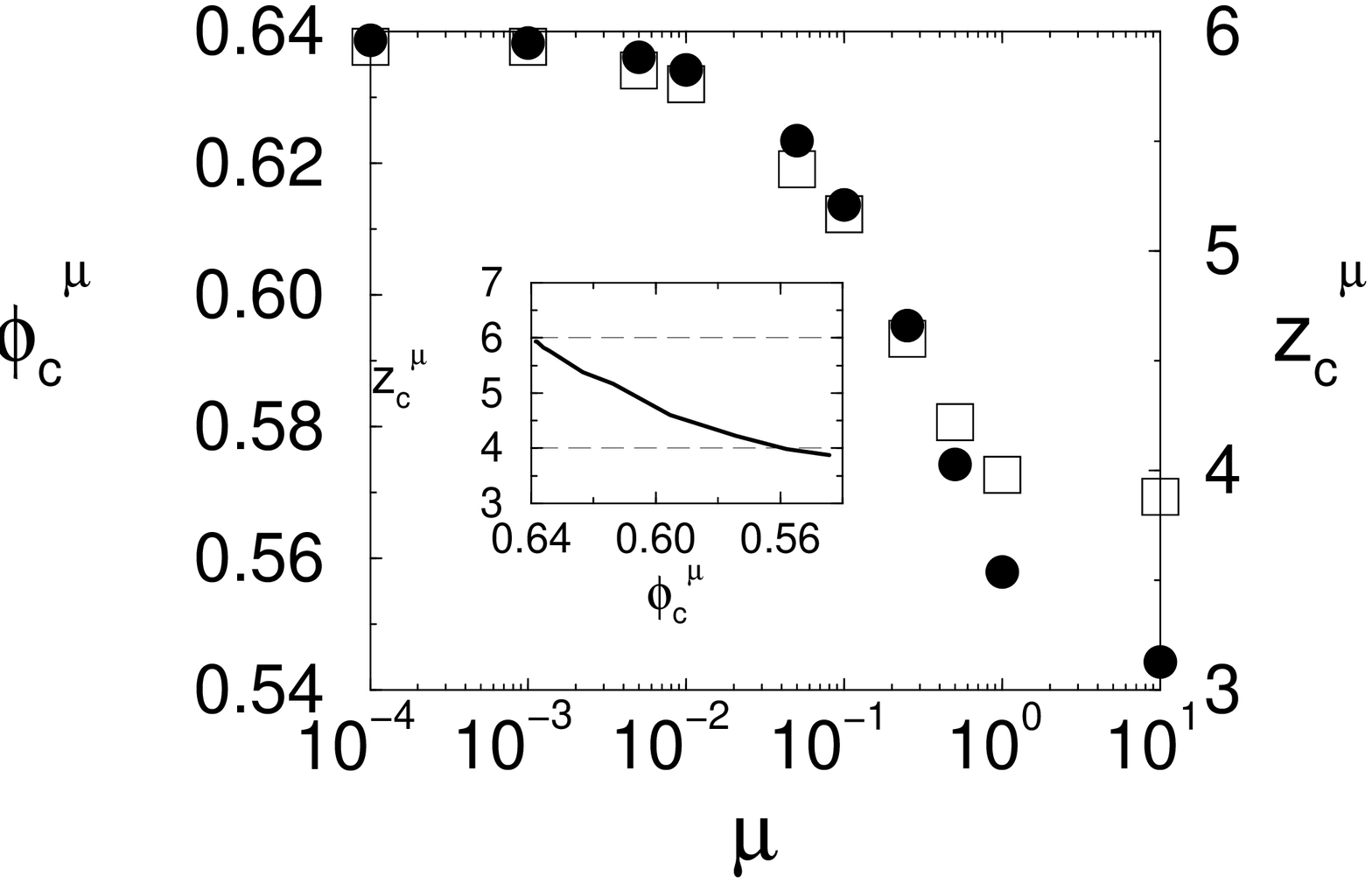}
  \includegraphics[width=7.5cm]{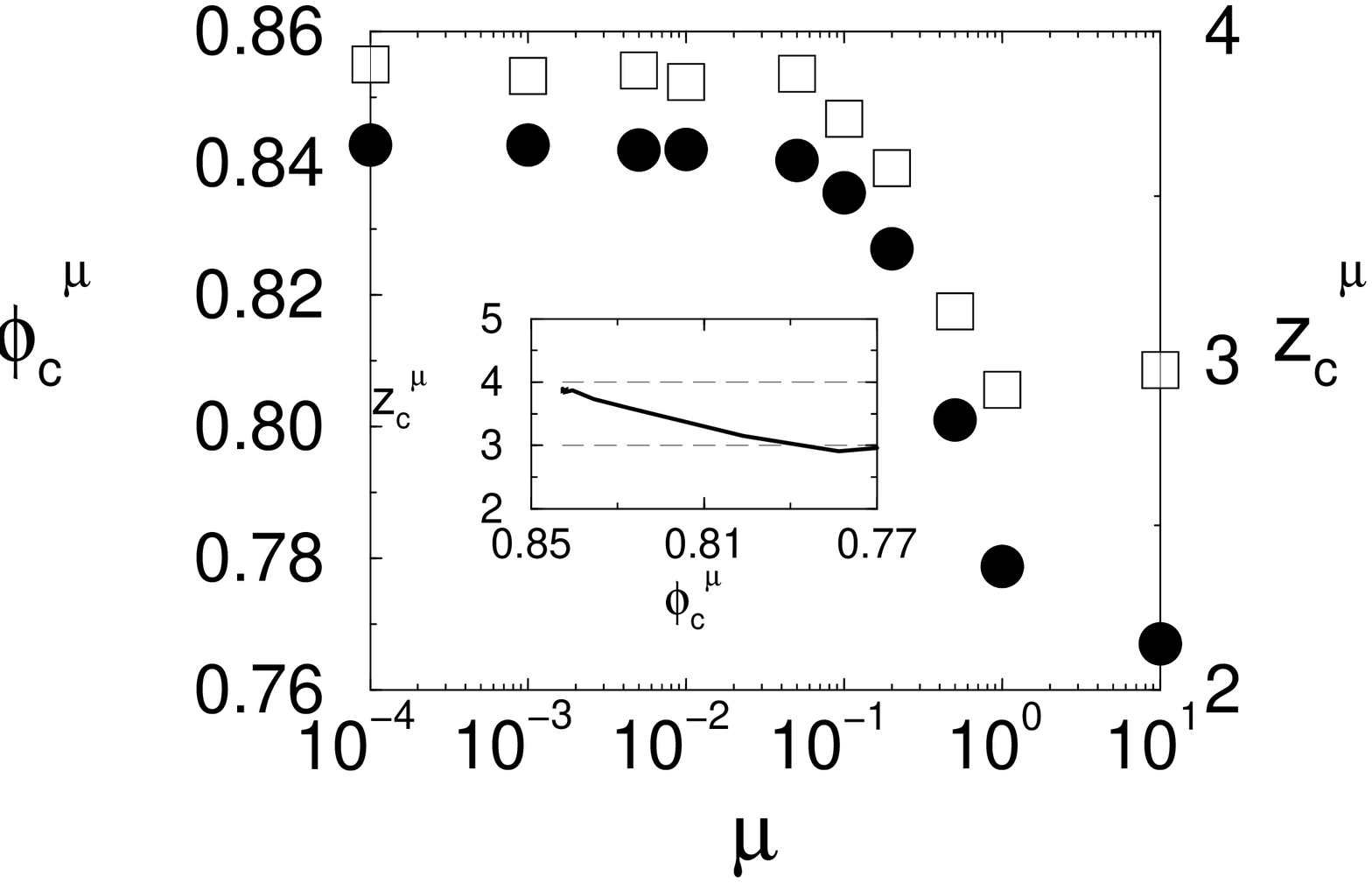}
  \caption{Dependence of the critical values of the packing fraction
    $\phi^{\mu}_{c}$ (filled circles), and coordination number $z^{\mu}_{c}$
    (open squares), on the particle friction coefficient $\mu$, for
    monodisperse spheres in $3D$ (upper panel) and bidisperse discs in $2D$
    (lower panel).  The insets are parametric plots of $\phi_{c}^{\mu}$
    against $z_{c}^{\mu}$. Symbol size is representative of sample-to-sample
    fluctuations and error bars.}
  \label{fig1}
\end{figure}

The question now arises, to what extent do these developing ideas of jamming
apply more generally in systems with non-central force laws, such as
frictional, granular packings?  The identification of the jamming threshold in
frictionless sphere packings to the well-known rcp state begs the question,
can these same ideas lead to a more concrete definition of the random loose
packing state for frictional spheres? These concerns are investigated here by
studying how interparticle friction affects the jamming properties of
monodisperse spheres. Here it is shown that the extrapolated values of the
packing fraction $\phi_{c}^{\mu}$, and the coordination number $z_{c}^{\mu}$,
at the jamming transition depend on the friction coefficient $\mu$, as
presented in \ref{table1} and \ref{fig1}, in agreement with conclusions of
recent experiments \cite{schroter3} and theory
\cite{makse1,leo9,makse2,wolf3,hecke7,makse7}, and that $\phi_{c} \rightarrow$
$\phi_{\rm rlp}$ and $z_{c} \rightarrow z^{\mu=\infty}_{\rm iso} = 4$ (in
$3D$, $=3$ in $2D$) in the limit of large friction. Here, $z^{\mu=\infty}_{\rm
  iso}$ represents the isostatic state for frictional spheres, corresponding
to the minimally required coordination number for infinitely hard, frictional
spheres \cite{edwards3}. Although the main focus of this study is three
dimensional packings, for completeness, some information for $2D$ packings is
also provided in \ref{fig1} and \ref{table1}.
\begin{table}[h]
  \caption {Extrapolated values of the packing fraction $\phi_{c}^{\mu}$, and
    coordination number $z_{c}^{\mu}$, at the jamming transition for different
    friction coefficient $\mu$ for $3D$ monodisperse spheres (top) and $2D$
    bidisperse discs (bottom). The associated errors on these estimates
    range from less than 1\% for small $\mu$ to 4\% at larger values of $\mu$.}
  \label{table1}
  \begin{tabular}{c}
  $3D$\\
  \end{tabular}\\
  \begin{tabular}{|c|c|c|c|c|c|c|c|c|}
    \hline
    $\mu$ &  0 & 0.001 & 0.01 & 0.1 & 0.2 & 0.5 & 1 & 10 \\\hline
    $\phi_{c}^{\mu}$ & 0.639 & 0.638 & 0.634 & 0.614 & 0.595 & 0.574 & 0.556 & 0.544 \\\hline
    $z_{c}^{\mu}$ & 5.96 & 5.93 & 5.76 & 5.17 & 4.60 & 4.22 & 3.98 & 3.88 \\\hline
  \end{tabular}\\
  \vspace{0.5cm}
  \begin{tabular}{c}
  $2D$\\
  \end{tabular}\\
  \begin{tabular}{|c|c|c|c|c|c|c|c|c|}
    \hline
    $\mu$ &  0 & 0.001 & 0.01 & 0.1 & 0.2 & 0.5 & 1 & 10 \\\hline
    $\phi_{c}^{\mu}$ & 0.843 & 0.843 & 0.842 & 0.0.836 & 0.827 & 0.801 & 0.779 & 0.767 \\\hline
    $z_{c}^{\mu}$ & 3.96 & 3.86 & 3.85 & 3.73 & 3.59 & 3.15 & 2.91 & 2.97 \\\hline
  \end{tabular}
\end{table}

To date, the influence of friction has been studied in the context of three
dimensional granular packings \cite{yu1,leo9,leo11,aste3}, and, to some
extent, their jamming properties \cite{makse1,makse2,makse7}. Makse and
co-workers \cite{makse7} recently made significant contributions to our
understanding of frictional packings from a theoretical point of view and the
results presented here are fully consistent with these previous studies. In
saying that, however, most earlier studies have not accurately addressed the
issue as to whether frictional packings jam in the same way as frictionless
ones nor has the issue of history dependence been seriously considered. These
are the two principal themes investigated in this study.

What is apparent is that the computational models employed are fully relevant
to address realistic systems. In particular, experiments
\cite{durian6,behringer9} and simulations \cite{nakanishi1,hecke7,hecke8} of
pseudo-$2D$, jammed, disc packings show good agreement suggesting that the
phenomenon of jamming seems to be applicable to real, frictional
materials. Here, it is shown that friction does indeed play an essential role
in determining the jamming transition for $3D$ sphere-packings as is the case
in two dimensional simulations of the Leiden group \cite{hecke7,hecke8}. The
principal result is the contention that the concept of the random loose
packing state becomes a friction-dependent property. The commonly quoted value
for $\phi_{\rm rlp} \approx 0.55$, is only realised in the limit of large
friction. Moreover, power law scaling is observed in the pressure and excess
coordination relative to the friction-dependent jamming transition and that
these critical values also exhibit power law behaviour with friction
coefficient.

\section*{Simulation Model}
The simulation technique used here employs $N=1024$ monodisperse, inelastic,
soft-spheres of diameter $d=1$ and mass $m=1$, within a cubic simulation box
with periodic boundary conditions without gravity. Particles interact only on
contact when they overlap, at which point they are considered to be contacting
neighbours, through a repulsive, linear spring-dashpot. The repulsive force is
characterised by the particle stiffness $k_{n,t}$, and inelasticity by the
coefficient of restitution $e_{n,t}$, in the normal ($n$) and tangential ($t$)
directions with respect to the contact surfaces. Unless otherwise stated, for
$\mu > 0$, a static friction law tracks the history of the friction forces
over the lifetime of a contact that satisfies the Coulomb yield criterion
\cite{cundall1,walton1,leo9,cupbook2}. In this work, the particle Poisson
ratio, $\nu = \frac{1 - \lambda}{4 + \lambda} = 0$\cite{rothenburg3}, where
the ratio of the tangential to the normal particle stiffness is denoted by,
$\lambda \equiv \frac{k_{t}}{k_{n}}=1$, for this study.

The initial configurations were generated by taking a dilute assembly of
particles in a disordered, liquid-like configuration, then instantaneously
quenching these configurations into over-compressed, jammed packed states at
$\phi_{i} = 0.65$. After this rapid compression the configurations were then
allowed to relax into a mechanically stable state. Unless otherwise stated,
during this initial compression friction was switched off. In an alternative
protocol to be discussed later, friction was switched on during the initial
quench to the over-compressed state.

The packing fraction of these mechanically stable packings were then
incrementally decreased towards the jamming threshold with the friction
coefficient set at the desired value. After each incremental change in the
packing fraction, the packing was rapidly quenched back into a mechanically
stable state by setting $e_{n,t} \approx 0$. This procedure was continued
until the difference in the potential energy between successive increments
$\lesssim 10^{-16}$, at which point the simulation run terminates. To improve
statistical uncertainty, all results are averaged over at least five
independent realisations.

\section*{Results and discussion}

The jamming transition is identified as the point at which the pressure $p$,
computed from the contact stresses, goes to zero. The jamming transition
packing fraction $\phi_{c}$, is obtained as a fitting parameter by
extrapolating the $\phi-p$ curve to $p=0$. Likewise, the coordination number
at the transition, $z_{c}$, is similarly obtained by fitting the $(p,z)$ data
to the functional form used for frictionless spheres \cite{ohern2} and bubbles
\cite{durian5}: $p \propto (z - z_{c})^{1/2}$.  This same power law exponent
has also been used to fit experimental \cite{behringer9} and numerical data
\cite{makse2,hecke7} of frictional systems. 
\begin{figure}[h]
(a)\hfil
  \includegraphics[width=2.3cm,height=2.3cm]{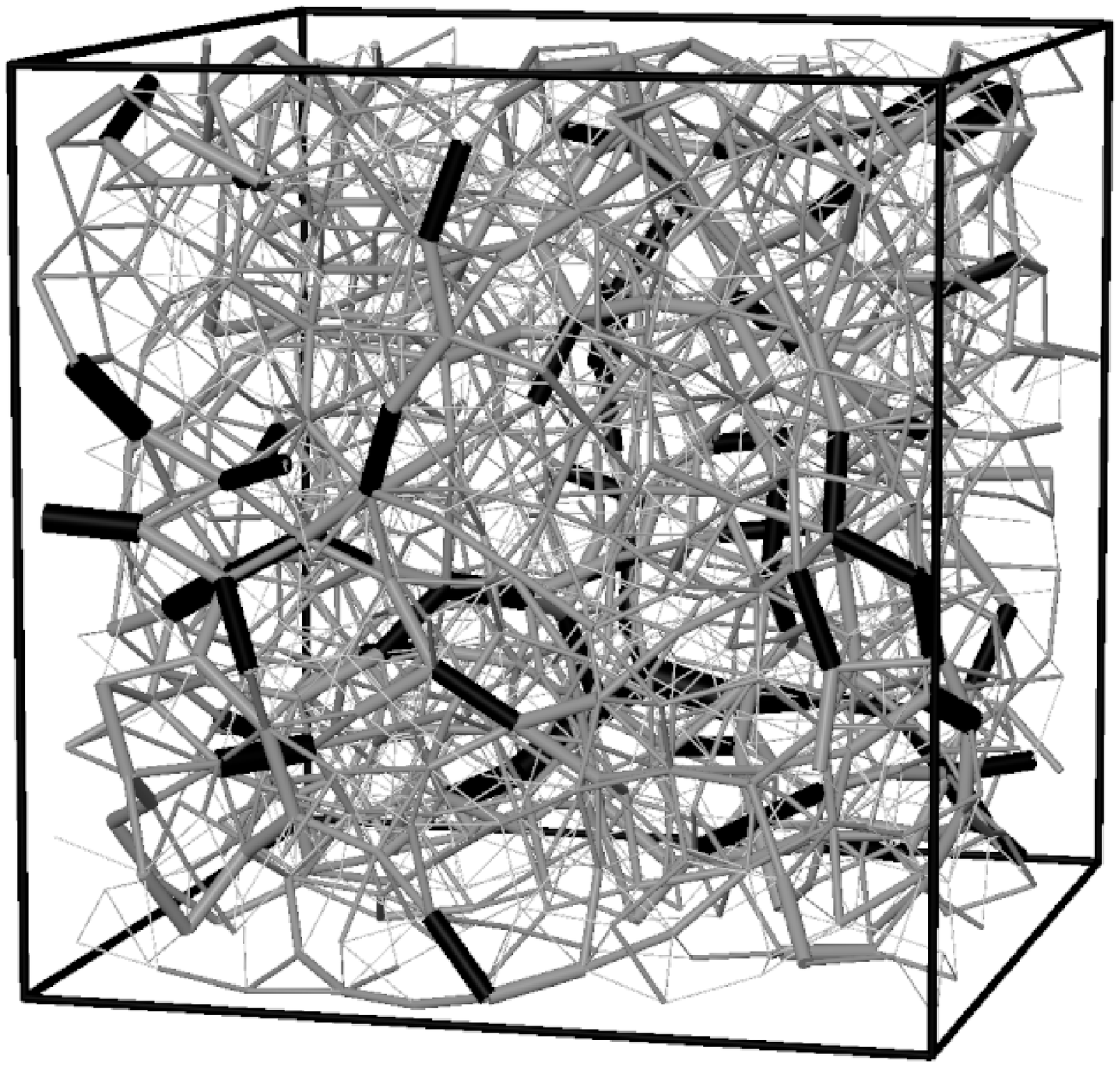}
(b)\hfil
  \includegraphics[width=2.3cm,height=2.3cm]{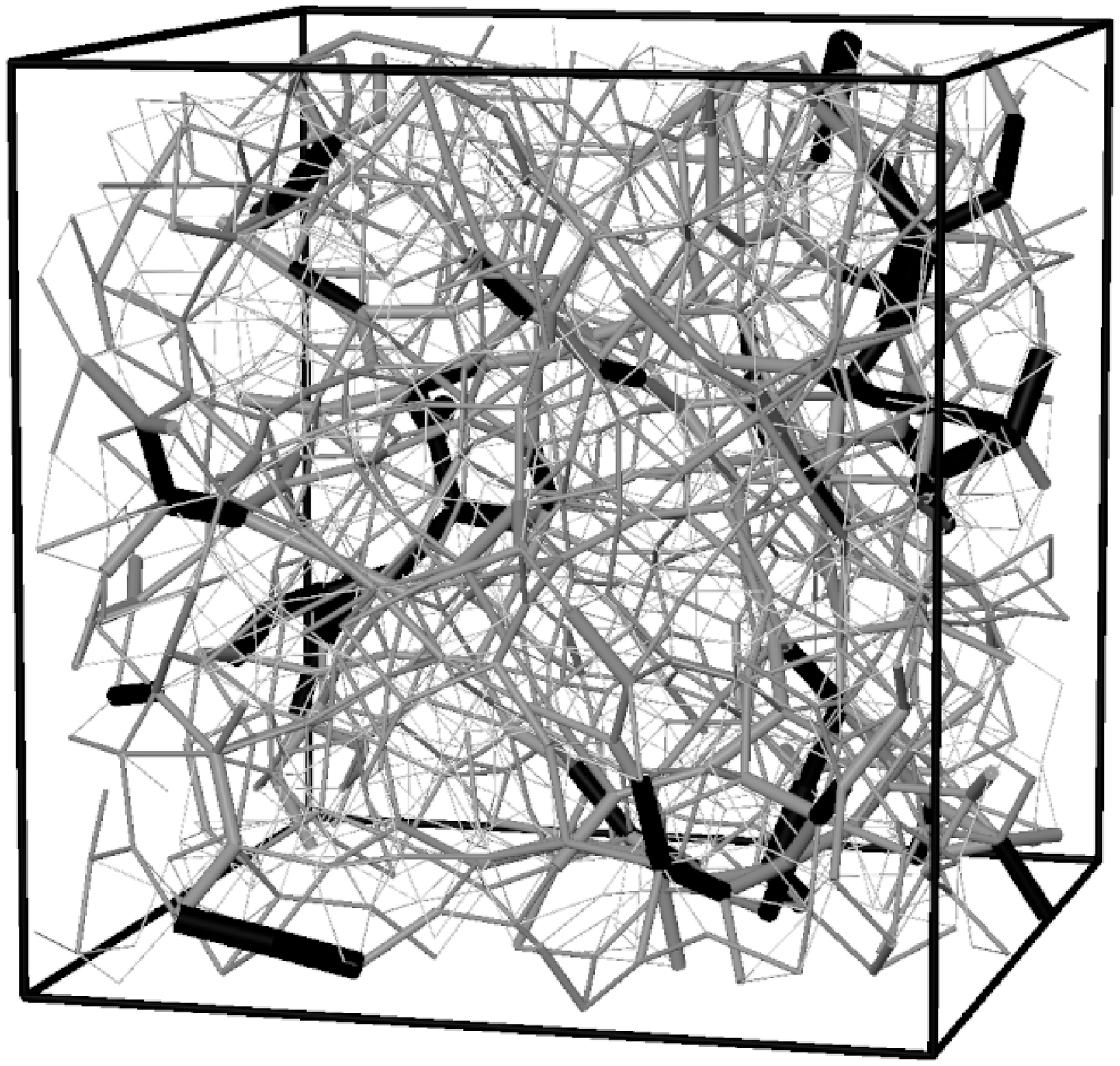}
(c)\hfil
  \includegraphics[width=2.3cm,height=2.3cm]{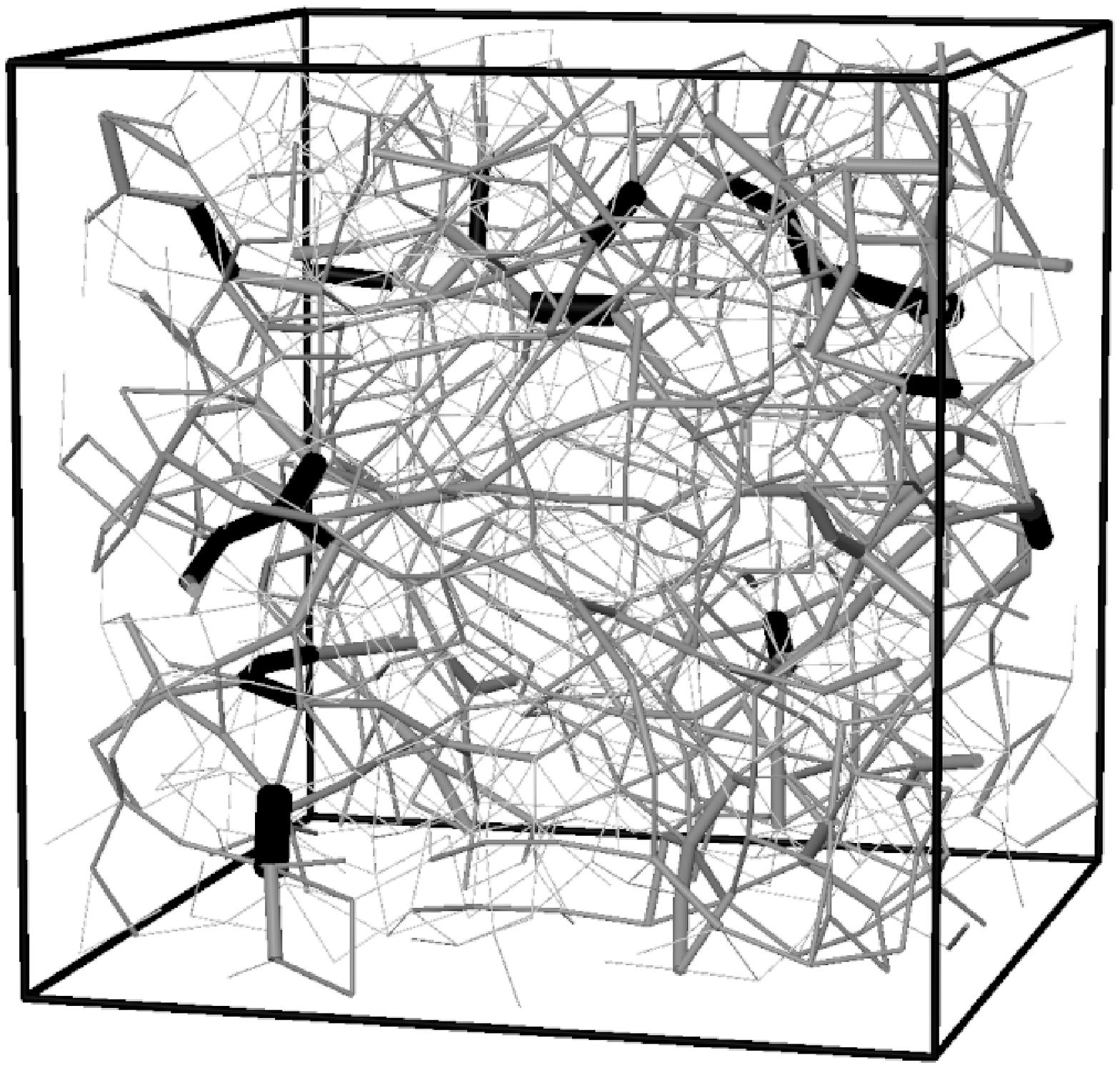}
  \caption{Snapshots of the normal force networks for three different friction
    coefficients at approximately the same pressures or equivalently the same
    $\Delta \phi^{\mu} \approx 10^{-4}$; (a) $\mu = 0$, (b) $\mu = 0.1$, and
    (c) $\mu = 1$. Lines represent the normal forces between particles in
    contact. Thicker, darker lines indicate larger forces. Particles not shown
    for clarity.}
  \label{fig2}
\end{figure}

\begin{figure}[h]
(a)\hfil
  \includegraphics[width=3.5cm,height=3.5cm]{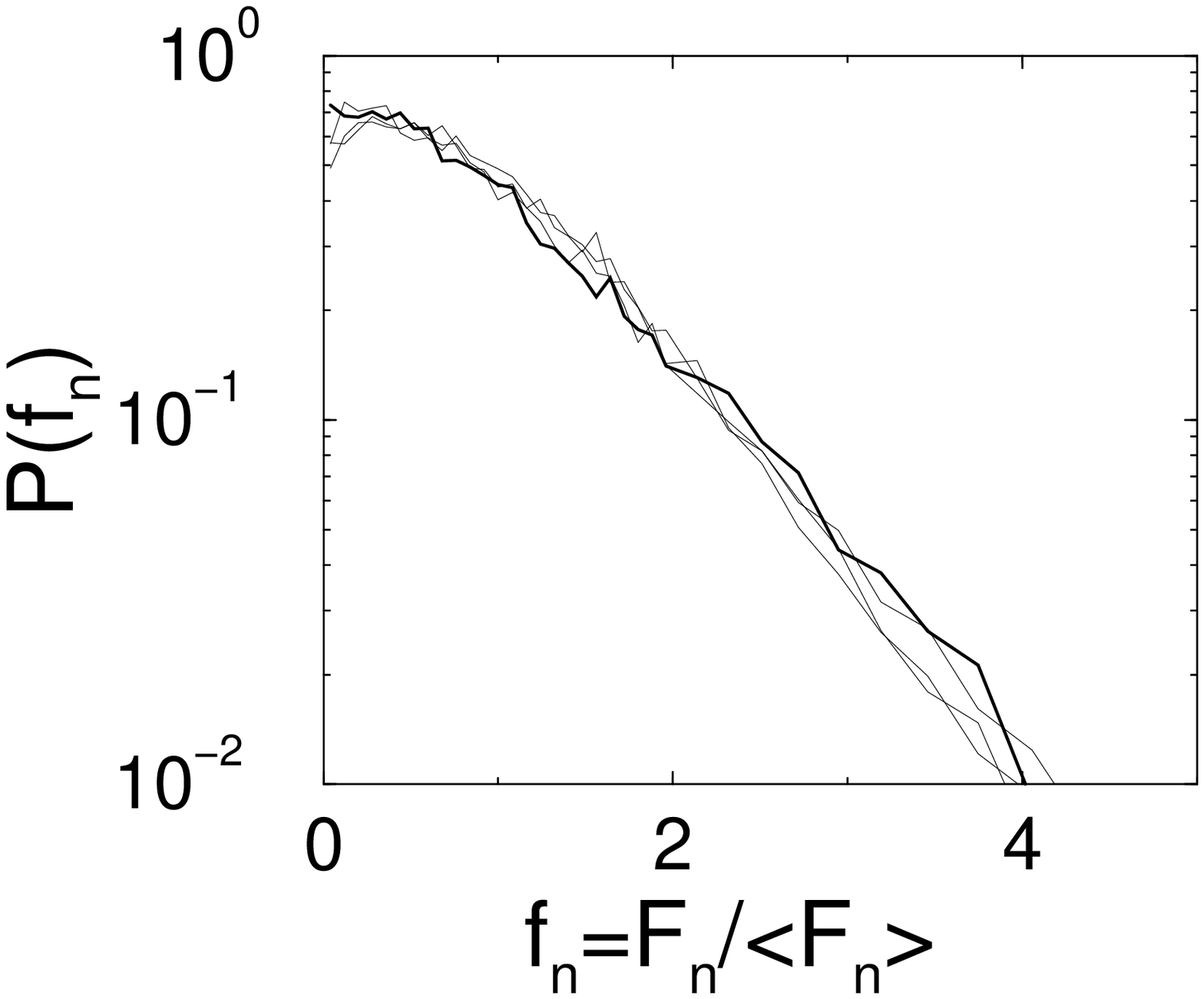}
(b)\hfil
  \includegraphics[width=3.5cm,height=3.5cm]{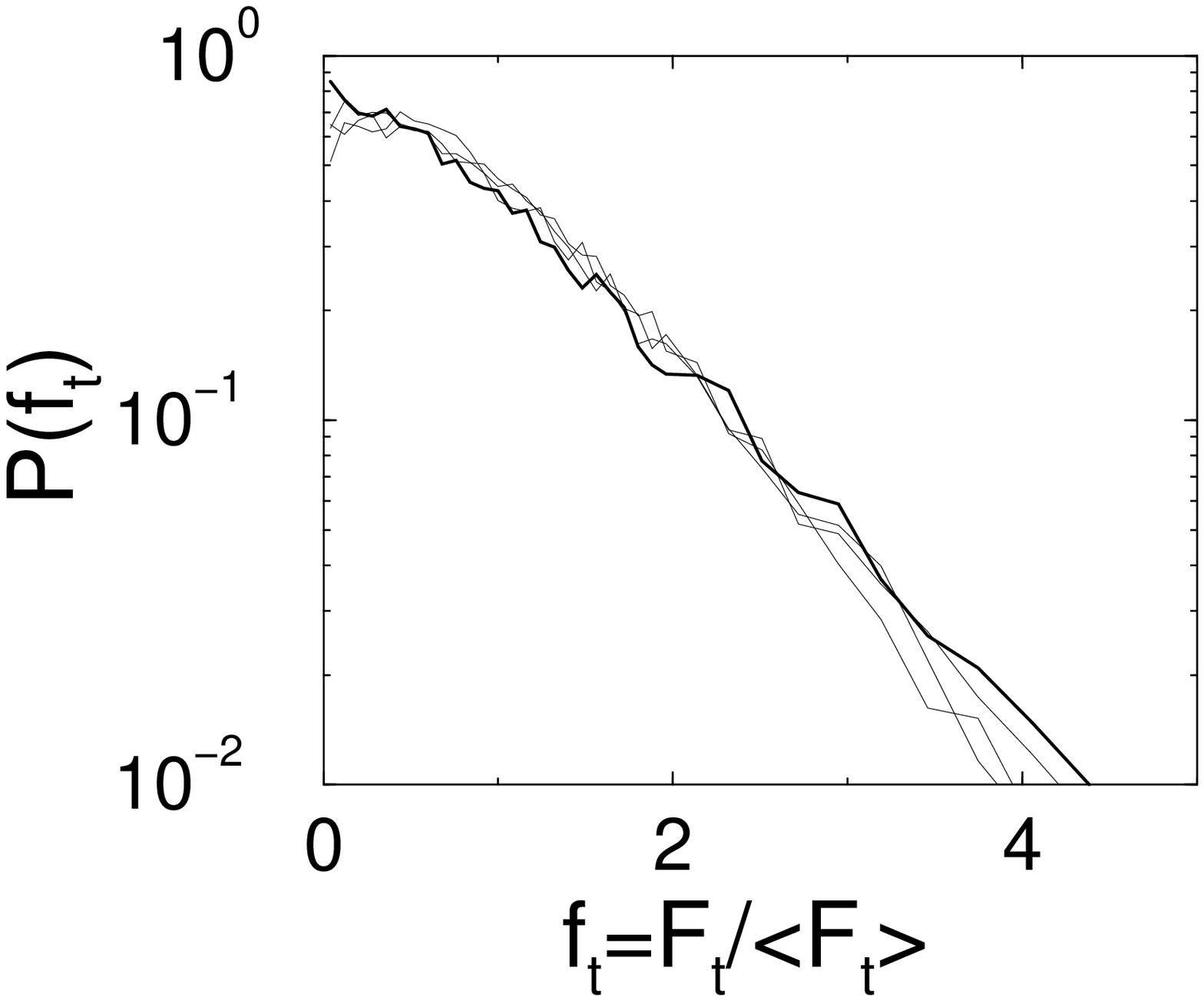}
  \caption{Distributions, $P(f_{n,t})$, of the, (a) normal $F_{n}$, and (b)
    tangential $F_{t}$, contact forces, normalised by their respective mean
    values. Data at approximately the same $\Delta \phi^{\mu} \approx 10^{-4}$
    for four different friction coefficients are shown: $\mu = 0.001, 0.1,
    0.25, 1$. The data at $\mu=0.25$ is emphasised.}
  \label{fig3}
\end{figure}

The transition values, $\phi_{c}^{\mu}$ and $z_{c}^{\mu}$, for different
$\mu$, are shown in \ref{table1} and plotted in \ref{fig1}. The insets in
\ref{fig1} represent the boundary between stable and unstable states. For
$3D$, monodisperse spheres with $\mu=0$, $\phi_{c}^{\mu=0} = 0.64 \pm 0.001$
and $z_{c}^{\mu=0} = 5.96 \pm 0.05$, are indistinguishable from previous
studies that used similar and different algorithms
\cite{makse2,ohern2,torquato5,leo9,roux3}. These results are consistent with
$\phi_{\rm rcp}$ and $z_{\rm iso}^{\mu=0}$. As $\mu$ increases,
$\{\phi_{c}^{\mu},z_{c}^{\mu}\}$ decrease. In the limit of large friction,
$\mu \geq 1$, these values saturate at about, $\phi_{c}^{\mu=\infty} \approx
0.55$ and $z_{c}^{\mu=\infty} \approx 4$, which coincide with the values of
$\phi_{\rm rlp}$ and $z_{\rm iso}^{\mu=\infty}$. To check that these values do
indeed correspond to the hard-particle limit, the particle stiffnesses were
varied over five orders of magnitude keeping their ratio $\lambda =1$. The
resulting critical values obtained were all indistinguishable within
statistical error. For completeness data for two dimensional, bidisperse disc
packings (with particle size ratio 1:1.4) is also shown in \ref{fig1} and
\ref{table1}. The $2D$ data are consistent with other jamming results
\cite{hecke7}, simulations of sheared granular materials \cite{rothenburg4},
and studies on force indeterminacy \cite{wolf3}.

This behaviour can be reasoned by the following arguments. For small friction
coefficients the tangential forces do not contribute significantly to the
stability of the packing and hence the packings are not that different from
frictionless systems. This behaviour persists, as shown in \ref{fig1}, up to
friction coefficients approaching order unity where the typical tangential
force first becomes comparable to the normal force. Hence, the friction forces
start to play a significant role in stabilising the packing. The precise value
of the friction coefficient where this transition occurs is likely to depend
on the particle Poisson ratio $\nu$. In the work presented here $\nu =
0$. Preliminary data with both negative and positive values of $\nu$ indicate
a possible dependence of the critical values on $\nu$, but these differences
are practically within statistical uncertainty. Other granular simulations
\cite{leo9} with $\nu=\frac{1}{6}$ find similar behaviour in
the packing fraction and coordination number indicating that the data
presented here is a general stability property of frictional systems.

Associated changes in the structural properties of the packings with friction
are highlighted in \ref{fig2} and \ref{fig3}, where force network information
is presented. \ref{fig2} shows the normal force networks where the normal
forces between particles in contact are represented by lines connecting the
centres of the particles. Thicker, darker line shading indicate forces with
increasing magnitude. As the critical packing fractions and coordination
numbers decrease with increasing friction then density of contacts likewise
decreases as is seen in these snapshots. It is also worth pointing out that
although the global force network by necessity pervades the system, there is
little indication of long-ranged, correlated, force chain structures in the
normal forces. \ref{fig3} shows the distributions of the magnitudes of the
normal and tangential forces for packings close to the critical states for
different friction coefficients. In all cases, the distributions are
characterised by exponential-like tails at larger forces. However, there are
some subtle changes occurring at the small force plateau-peak region. Notably,
on careful inspection of the data at $\mu=0.25$, which has been emphasised in
the figures and corresponds to a value of the friction coefficient the
smallest forces only for friction coefficients where the critical values are
decreasing with friction in \ref{fig1}, both the normal and tangential
distributions exhibit an upturn at the smallest forces. Thus indicating that
the decrease in the critical values indicated in \ref{fig1} is correlated with
the fraction of small forces in the system.

\subsection*{Scaling}

\begin{figure}[h]
  \includegraphics[width=8cm]{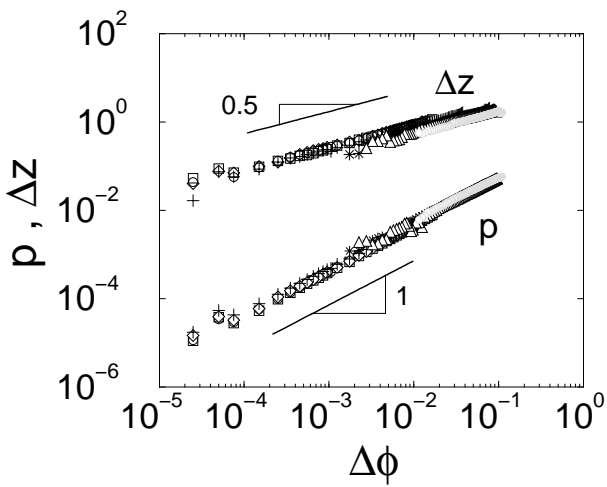}
  \caption{Power law scaling of the pressure $p$, and excess coordination
    number $\Delta z$, as a function of the excess packing fraction,
    $\Delta\phi$, over and above the extrapolated values at the jamming
    threshold \{$z^{\mu}_{c},\phi^{\mu}_{c}\}$. Different symbols represent
    different friction coefficients: $\mu = $ 0 ($\Diamond$), 0.001 ($\circ$),
    0.01 ($\square$), 0.1 ($+$), 0.5 ($*$), 1 ($\triangle$), and 10 (grey
    $\circ$). The solid lines represent power laws with exponents 0.5 and 1 as
    indicated.}
  \label{fig4}
\end{figure}

Motivated by earlier studies on frictionless systems, it is possible to
rescale all the data for different $\mu$, by using the measure $\Delta\phi
\equiv \phi - \phi_{c}^{\mu}$, as the distance for each system for a given
$\mu$, from their respective jamming transitions, i.e. it is now not
appropriate to measure the distance to the jamming transition using just the
zero-friction value $\phi_{c}^{\mu=0}$, for different $\mu$. \ref{fig4} shows
the power-law scaling, data collapse for the pressure and the excess
coordination number $\Delta z \equiv z - z_{c}^{\mu}$, for all friction
coefficients:
\begin{equation}
\begin{array}{cc}
P \propto \Delta\phi,\\
\Delta z \propto (\Delta\phi)^{1/2}.
\end{array}
\label{eq1}
\end{equation}
This result shows that jamming frictional spheres look like jamming
frictionless spheres provided one identifies the friction-specific jamming
threshold values $\phi_{\rm c}^{\mu}$ and $z_{\rm c}^{\mu}$.

\begin{figure}[h]
(a)\hfil
  \includegraphics[width=3.5cm,height=3.5cm]{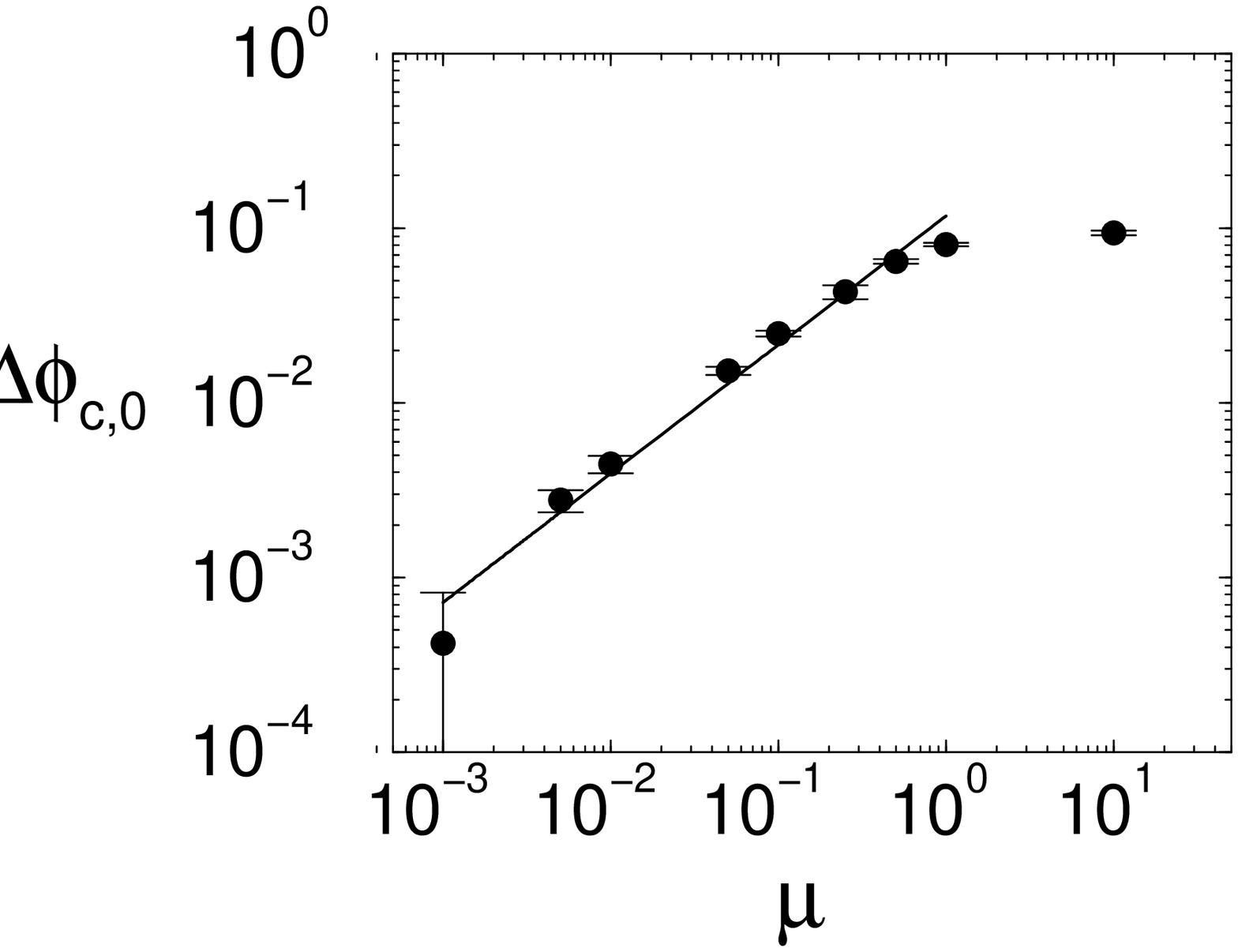}\hfil
(b)\hfil
  \includegraphics[width=3.5cm,height=3.5cm]{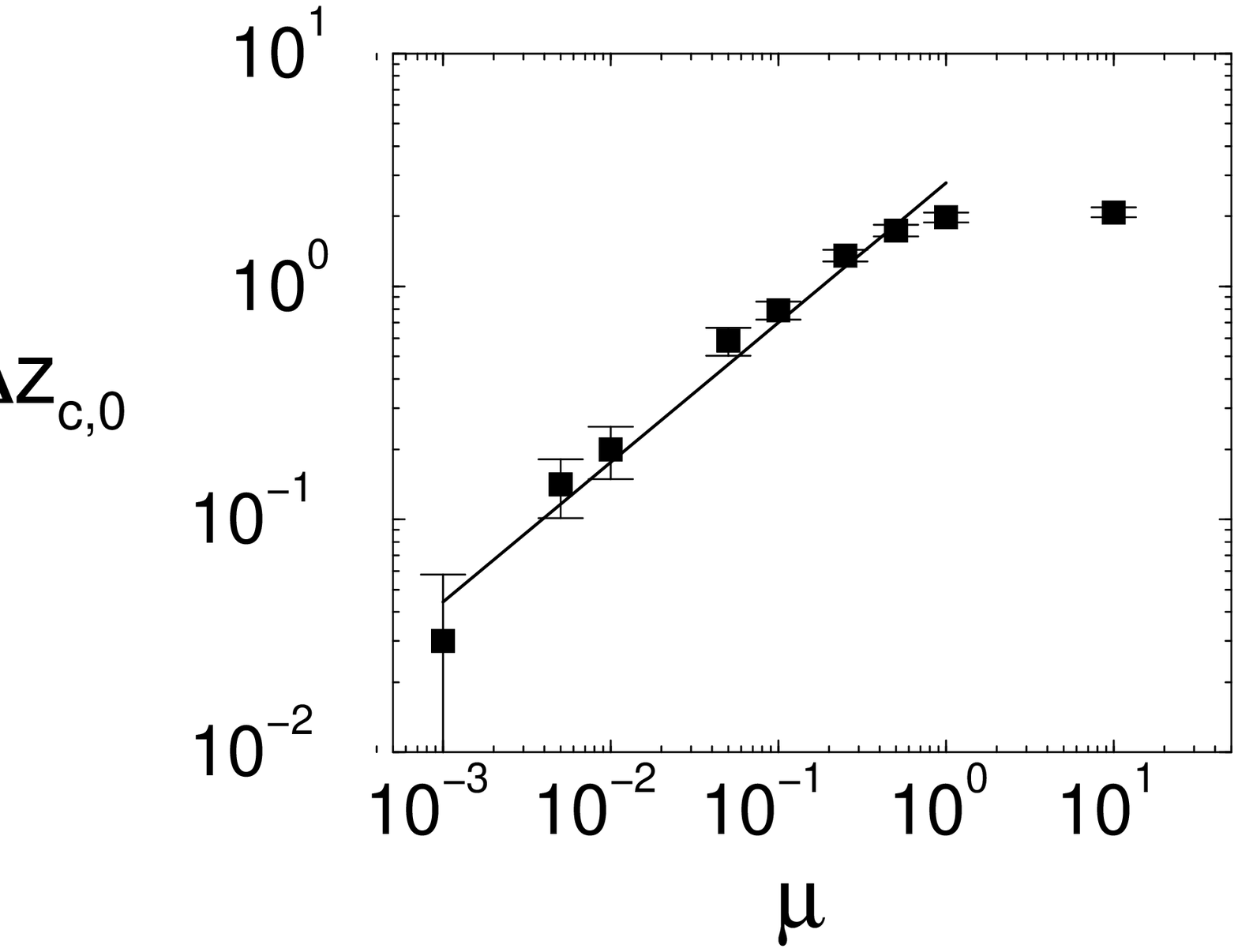}
  \caption{Power laws for, $\Delta\phi_{\rm c,0} \equiv (\phi_{c}^{\mu=0} -
    \phi_{c})$ ($\bullet$), the packing fraction, and $\Delta z_{c,0} \equiv
    (z_{c}^{\mu=0} - z_{c})$ ($\blacksquare$), the coordination number, at
    jamming relative to the zero friction values, over several orders of
    magnitude in $\mu$. Solid lines are power law fits to data described in
    Eq.~\ref{eq2}, with power law exponents $0.74\pm 0.09$ and $0.64\pm 0.04$
    in (a) and (b) respectively.}
  \label{fig5}
\end{figure}

\begin{figure}[!]
(a)\hfil
  \includegraphics[width=3.5cm,height=3.5cm]{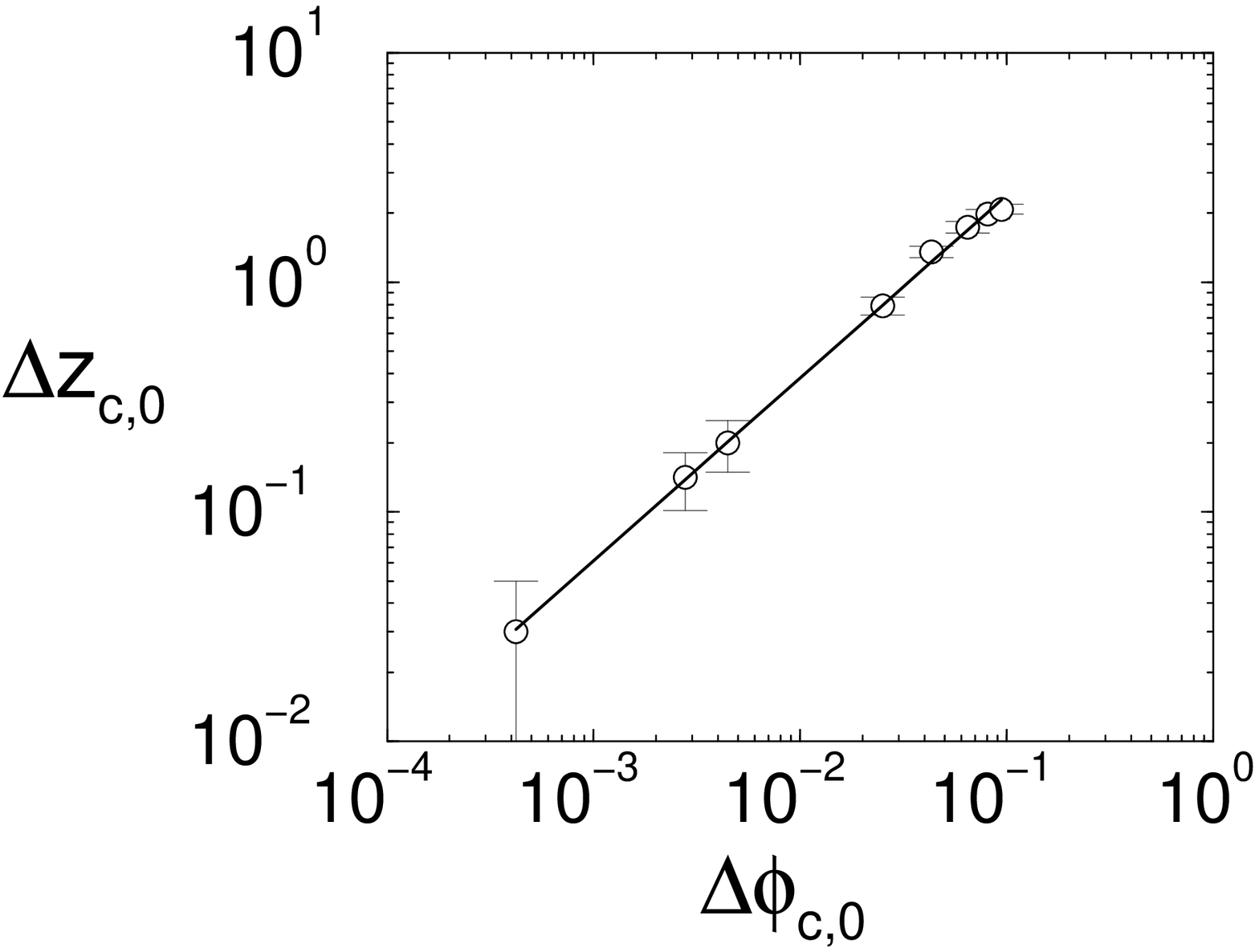}\hfil
(b)\hfil
  \includegraphics[width=3.5cm,height=3.5cm]{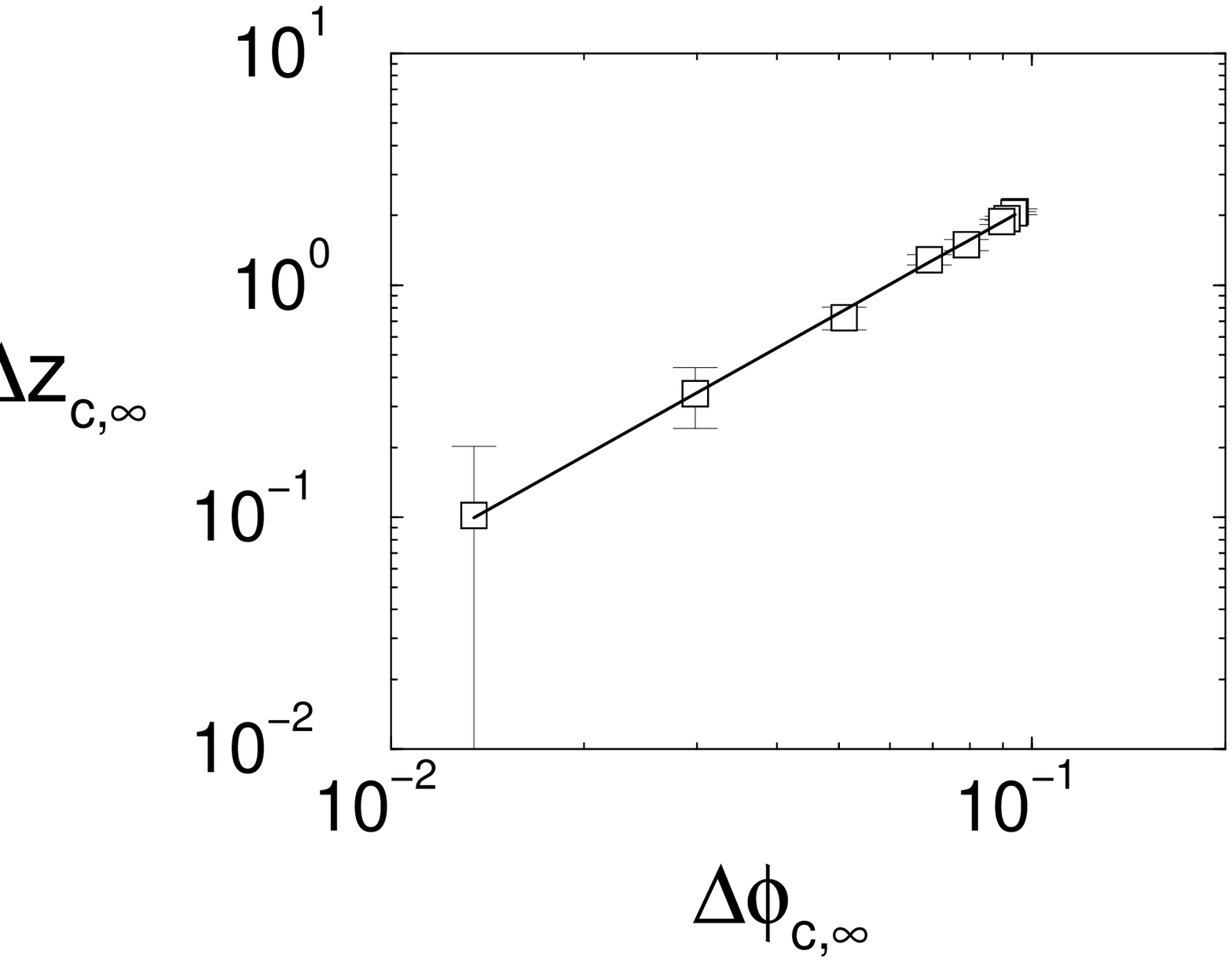}
  \caption{Power laws between; (a) $\Delta\phi_{c,0} \equiv (\phi_{c}^{\mu=0}
    - \phi_{c})$, and $\Delta z_{c,0} \equiv (z_{c}^{\mu=0} - z_{c})$, the
    jamming values relative to the zero-friction state, and (b)
    $\Delta\phi_{c,\infty} \equiv (\phi_{c} - \phi_{c}^{\mu=\infty})$ and
    $\Delta z_{c,\infty} \equiv (z_{c} - z_{c}^{\mu=\infty})$, the jamming
    values relative to the large-friction state. Solid lines are power law
    fits to data described by Eq.~\ref{eq3}, with power law exponents $0.82\pm
    0.06$ and $1.54\pm 0.08$ in (a) and (b) respectively.}
  \label{fig6}
\end{figure}

Furthermore, the jamming thresholds for the packing fractions and coordination
numbers exhibit several power-law relationships as a function of the friction
coefficient, as shown in \ref{fig5} and \ref{fig6}. Up to the large-friction
limit, where both $\phi_{\rm c}^{\mu}$ and $z_{\rm c}^{\mu}$ practically
saturate, the transition values relative to the $\mu=0$ values, scale as,
\begin{equation}
  \begin{array}{cc}
    \Delta\phi_{\rm c,0} \equiv (\phi_{\rm c}^{\mu=0} - \phi_{\rm c}^{\mu})
\sim \mu^{0.74[9]},\\
    \Delta z_{\rm c,0} \equiv (z_{\rm c}^{\mu=0} - z_{\rm c}^{\mu}) \sim
\mu^{0.64[4]}.
  \end{array}
  \label{eq2}
\end{equation}

The corresponding measures, $\Delta\phi_{\rm c,\infty} \equiv
(\phi_{\rm c}^{\mu} - \phi_{\rm c}^{\mu=\infty})$ and $\Delta z_{\rm
  c,\infty} \equiv (z_{\rm c}^{\mu} - z_{\rm c}^{\mu=\infty})$,
relative to the infinite friction limit do not exhibit similar
relations, however, the following power laws are also observed:
\begin{equation}
  \begin{array}{c}
    \Delta\phi_{\rm c,0} \sim \left(\Delta z_{\rm
c,0}\right)^{1.23[10]},\\
    \Delta\phi_{\rm c,\infty} \sim \left(\Delta z_{\rm
c,\infty}\right)^{0.65[8]},\\
  \end{array}
  \label{eq3}
\end{equation}
It is not clear as to the origin of Eq.~\ref{eq2}, although the first relation
in Eq.~\ref{eq2} is consistent with Eq.~\ref{eq1}. These results are similar
to two dimensional systems \cite{hecke7}, where the corresponding four power
law exponents were obtained: $0.77$, $0.70$, $1.1$, and $0.59$, and thus hints
at some possible underlying universal properties of frictional packings
irrespective of dimensionality.

The results presented thus far have implications regarding the statistical
mechanical ensemble formulation of granular (random) packings espoused by
Edwards \cite{edwards5} and has been extensively explored by Makse and
co-workers \cite{makse7} and others \cite{henkes3}. Edwards revision of
statistical mechanics proposes that the properties of a granular packing can
be computed from a statistical average over equally probable
configurations. For infinitely hard and frictional particles the partition
function entering the analogue of the canonical ensemble is summed over the
full range of all possible states compatible with mechanical stability, from
random loose packing, $\phi_{\rm rlp} = 0.55$, up to $\phi_{\rm rcp} = 0.64$
\cite{nowak1}. However, the results presented here suggest that generalising
to finite friction requires one to take into account friction-dependent
constraints: the sum-over-states should only include those mechanically stable
states compatible with the value of $\mu$, i.e. project onto the sum of states
from all possible configurations only the mechanically stable ones
\cite{nicodemi5}. Practically, this may be achieved by cutting-off the
sum-over-states at the appropriate value of $\phi_{c}^{\mu}$ \cite{schroter1}:
$\int_{\phi_{\rm rlp}}^{\phi} d\phi' \rightarrow \int_{\phi_{\rm
    c}^{\mu}}^{\phi} d\phi'$; or in the microcanonical formalism, states with
$z < z_{c}^{\mu}$ should be given zero weight. This is precisely the approach
taken by Makse and co-workers who deduced a frictional packing phase diagram
\cite{makse7} which is consistent with the results presented in \ref{fig1}.

\subsection*{Protocol Dependence}

The simulation protocol implemented here generates sphere packings that are
statistically identical to other works
\cite{ohern3,makse2,wolf3,hecke7,makse7}, resulting in similar values for the
transition packing fraction $\phi^{\mu}_{c}$, and coordination number
$z^{\mu}_{c}$, as presented in \ref{fig1} and \ref{table1}. These values
correspond to the hard sphere limit where the particles are just touching at
the point of jamming. This can be seen by comparing the various interaction
and simulation models (linear-spring or Hertzian interaction forces, molecular
dynamics or contact dynamics simulations)
\cite{leo9,makse7,hecke7,wolf3,roux3}. However, the path taken to arrive at
the jamming transition may differ, particularly when friction is
present. Because of the hysteretic nature of the frictional forces the precise
force configurations explored during this path will depend on the preparation
history of the packings.

\begin{figure}[!]
\includegraphics[width=8cm]{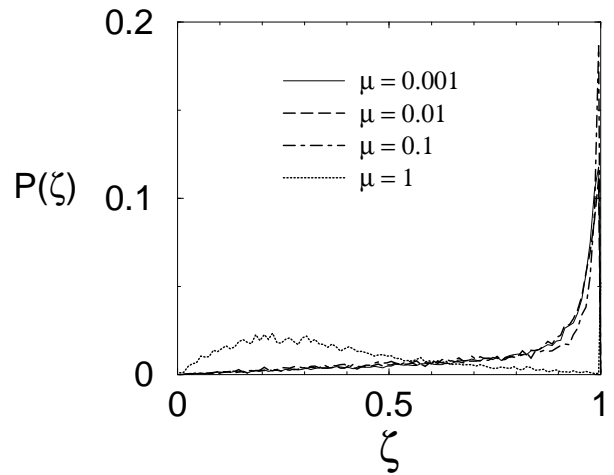}
\caption{Distributions, $P(\zeta)$, of the plasticity index, $\zeta \equiv
  F_{t}/\mu F_{n}$, where $F_{n,t}$ are the magnitudes of the normal and
  tangential forces, for different friction coefficients $\mu$, within
  $10^{-4}$ of the jamming threshold.}
  \label{fig7}
\end{figure}
The major factor influencing the frictional forces are the fraction of
slipping contacts $n_{s}$. The plasticity index is defined by \cite{leo9},
$\zeta \equiv F_{t}/\mu F_{n}$, where $F_{n,t}$ are the magnitudes of the
$\{$normal,tangential$\}$ forces. Here, slipping contacts are identified as
those contacts at or on the verge of yielding: $\zeta \geq 0.95$. The
qualitative nature of the results are not sensitive to the precise definition
of slipping contacts. The data of \ref{fig7} shows how the probability
distributions $P(\zeta)$ depend of friction coefficient for packings prepared
very close to the jamming threshold. For small friction coefficients $\mu <
0.2$, the distributions are dominated by a large fraction of contacts that are
close to yielding, $\zeta>0.9$. These systems correspond to points in
\ref{fig1} just before the rapid decrease in $\phi^{\mu}_{c}$ with
$\mu$. However, for larger friction coefficients, $\mu > 0.5$, where the
transition values begin to saturate, the weight of the distribution shifts to
smaller $\zeta$. Hence, a majority of contacts become stabilised as seen by
the growth of a hump at $\zeta \approx 0.25$ for $\mu=1$ in
\ref{fig7}. Similar changes in $P(\zeta)$ have been seen in simulations of
confined packings \cite{landry1} and two dimensional systems \cite{hecke7}.

\begin{figure}[!]
\includegraphics[width=8cm]{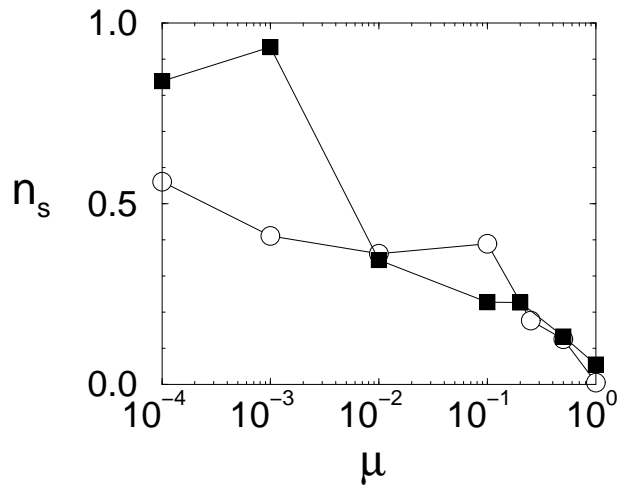}
\caption{Fraction of slipping contacts $n_{s}$ at $\Delta\phi(\mu) \approx
  10^{-3} - 10^{-4}$ over a large range in friction coefficients. By
  definition, $n_{s}(\mu=0) = 1$. The two sets of data represent two slightly
  different protocols as discussed in the text. Differences at small friction
  values diminish at larger friction coefficients for the two protocols.}
  \label{fig8}
\end{figure}
Although the qualitative features of the distributions shown in \ref{fig7} are
robust over different protocols, the actual numbers of slipping contacts
depends on the preparation history of the packings. To illustrate this point
\ref{fig8} shows the fraction, $n_{s}$, of slipping contacts over a wide
range of friction coefficients for packings that are close to the jamming
threshold. The two sets of data correspond to slightly different packing
preparation histories. The open symbols are for the protocol used throughout
this work to generate the previously shown data: a dilute packing was
over-compressed with friction initially set to zero prior to the decompression
procedure. The filled symbols, on the other hand, are for packings where
friction was set to the desired value during the initial over-compression
stage. The differences between the two protocols clearly show up at the lower
friction coefficients where the newly modified protocol exhibits a much larger
fraction of slipping contacts. The rapid compression with friction leads to a
build up of the frictional forces that quickly causes contacts to reach the
Coulomb yield criterion as the system is decompressed towards the jamming
threshold. However, these differences diminish for larger friction
coefficients. The exact nature of the frictional build up and relaxation is
currently under investigation. It is also worth pointing out that the
extrapolated jamming thresholds are identical within statistical uncertainty
between the two protocols. Thus, the protocol dependence does not seem to
affect the geometrical properties of the packing, but will likely have an
effect on mechanical properties, such as yielding under flow or the
application of external forces.

\begin{figure}[!]
\includegraphics[width=8cm]{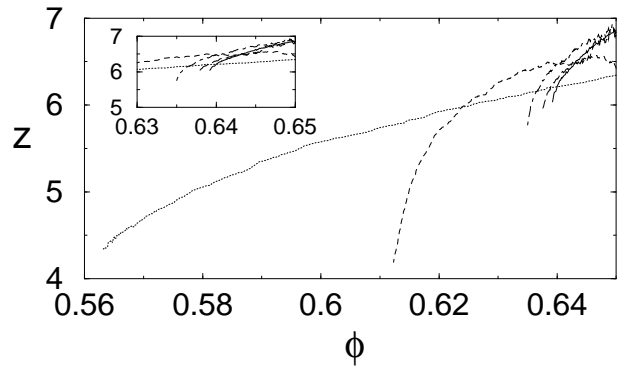}
\caption{Evolution of the coordination number, $z$, with decreasing
  packing fraction, $\phi$. Line styles represent different friction
  models: frictionless ($\mu = 0$, solid line); history-dependent
  static friction model (dotted), static-friction model reset at
  different rates (dot-dash and dash), and dynamic instantaneous
  friction model with no contact history (long-dash), $\mu = 1.0$.
  Inset shows the random close packing region.}
  \label{fig9}
\end{figure}

To probe the nature of protocol dependence further, the algorithm used to
generate the packings was also modified in the following ways: In one case a
dynamic friction model was used that depends only on the instantaneous
friction force. For this system the jamming threshold lies much closer to the
frictionless limit. In the other data, the history of the contacts was reset
at different intervals during the simulation procedure thereby suppressing the
build-up of frictional forces. The corresponding evolution of the coordination
number with packing fraction is shown in \ref{fig9} for packings these
packings where the contact history and frictional forces were treated
differently.

Thus, erasing the history of the frictional forces between particles in
contact can have a profound effect on the resulting evolution of the
packing. What \ref{fig9} demonstrates is that friction strongly influences the
jamming thresholds and hence the possible range of mechanically stable states
for a given friction coefficient. This may therefore explain, in part, the
observation why real, frictional granular materials, more often than not, tend
to form packings that are intermediate between the random close and random
loose limits: even minor rearrangements that cause the particles to
momentarily come out of contact will erase the history of frictional contacts
and thus the system will appear to be composed of particles of a lower
friction coefficient that their actual values. This aspect of the history
dependence of the frictional forces has been utilised recently in experiments
on real, frictional, granular particles made to mimic frictionless systems
\cite{ohern13}.

\section*{Conclusions}

In summary, cohesionless, frictional spheres that interact only on contact,
exhibit a jamming transition that shares many similarities with that for
frictionless systems, provided one identifies the appropriate
friction-dependent jamming transition packing fraction $\phi_{c}^{\mu}$, and
coordination number $z_{c}^{\mu}$. Using these friction-dependent quantities
rather than just the frictionless and frictional isostatic values may lead to
a better understanding of existing scaling results \cite{hecke8,makse11}. In
saying that, the role of packing generation protocol and friction-induced
history dependence persists as a topic that has received relatively little
attention to date. Understanding how packings traverse the packing stability
diagram of \ref{fig1} will likely help resolve some of these issues.

Random loose packing, $\phi_{\rm rlp} \approx 0.55$, appears to be a state
that can only be achieved in the large-friction limit. These results, however,
do not preclude the fact that even lower values of $\phi_{\rm rlp}$ may be
obtained using different protocols, such as sedimenting particles in a fluid
\cite{menon2}, or tuning the particle interactions to include additional
forces, such as cohesion \cite{yu8,ohern8}.

The power-law dependence on $\mu$ of both $\phi_{\rm c}$ and $z_{\rm c}$,
relative to the frictionless system hints at a more universal behaviour. The
origin of these scalings remain unclear although one possible explanation for
the results of \ref{fig5} and \ref{fig6} might lie in spatial correlations
between the frictional forces. As shown in \ref{fig8}, there is a systematic
decrease in the fraction of slipping contacts with increasing friction. One
might therefore expect that for small friction coefficients there are large
spatial regions of contacts that are slipping, whereas for large friction,
these are replaced by non-mobilised contacts. Hence, fluctuations are small in
either case. However, intermediate between these two extreme states,
large-scale, fluctuating regions might occur. Preliminary data suggests that
there is indeed an increase in correlations that follow these arguments and
such ideas are currently being pursued.

Applications of the results presented here could potentially pave the way
towards a generalised formulation of an equation of state for granular
materials. Progress along these lines has been made recently
\cite{makse7}. However, the hysteretic nature of the frictional forces can
have a significant effect on the properties of the packing. It is interesting
to note that the friction coefficients of many materials lie in the range
$0.01 \leq \mu \leq 1$. Coincidentally, these values of $\mu$ correspond to
the region of mechanically stable states where small changes in $\mu$ can lead
to large changes in the packing fraction and coordination number of the
packing. Hence, any processes that cause changes in the surface friction of
the constituent particles, such as roughening or polishing, can result in
packings with dramatically different stability properties. Alternatively,
materials could be designed with varying mechanical properties based on their
frictional properties.

\section*{Acknowledgements}

It is a pleasure to thank M. van Hecke and M. Schr\"oter for several
insightful discussions related to this work. Support from an SIU ORDA faculty
seed grant and the National Science Foundation CBET-0828359 is greatly
appreciated.

\providecommand{\url}[1]{\texttt{#1}}

\end{document}